\pdfoutput=1
\documentclass[aps,pre, twocolumn, groupedaddress]{revtex4-1}

\usepackage{lipsum}
\usepackage{float}
\usepackage{mathtools}
\usepackage{graphicx}
\usepackage{subfigure}
\usepackage{caption}
\usepackage{refstyle}
\usepackage{dcolumn}
\usepackage{amsmath}    
\usepackage{amssymb}
\usepackage{bm}
\usepackage{hyperref}
\usepackage{latexsym}
\usepackage{verbatim}
\usepackage[normalem]{ulem}
\usepackage{color}
\usepackage{parskip}
\def\beq{\begin{equation}}
\def\eeq{\end{equation}}

\def\beq{\begin{equation}}                          
\def\eeq{\end{equation}}                          
\def\bea{\begin{eqnarray}}                          
\def\eea{\end{eqnarray}}

\DeclareRobustCommand{\uvec}[1]{{%
  \ifcsname uvec#1\endcsname
     \csname uvec#1\endcsname
   \else
    \bm{\hat{\mathbf{#1}}}%
   \fi
}}
\preprint{}
\begin{document}
\preprint{}
\title{Ordering through learning in two-dimensional Ising spins}
\author{Pranay Bimal Sampat$^{1,2}$}
\email{pranayb.sampat.phy16@itbhu.ac.in}
\author{Ananya Verma$^{1}$}
\email{ananyaverma.phy18@itbhu.ac.in}

\author{Riya Gupta$^{1}$}
\email{riyagupta.phy18@itbhu.ac.in}
\author{Shradha Mishra$^{1}$}
\email{smishra.phy@iitbhu.ac.in}
\affiliation{$^{1}$Department of Physics, Indian Institute of Technology (BHU), Varanasi, India 221005}
\affiliation{$^{2}$Department of Physics, Brown University, Providence, Rhode Island 02912, USA.}

\date{\today}

\begin{abstract}
We study two-dimensional Ising spins, evolving through reinforcement learning using their state, action, and reward. The state of a spin is defined as whether 
it is in the majority or minority with its nearest neighbours. The spin updates its state using an $\epsilon$-greedy algorithm. The parameter $\epsilon$ plays the role equivalent to the temperature in the Ising model. We find a phase transition from long-ranged ordered to a disordered state as we tune $\epsilon$ from small to large values. In analogy with the  phase transition in the Ising model,	we calculate the critical $\epsilon$ and the three critical exponents $\beta$, $\gamma$, $\nu$ of magnetisation, susceptibility, and correlation length, 	respectively. A hyper-scaling relation $d \nu = 2 \beta + \gamma$ is obtained between the three exponents. The system is studied for different learning rates. The exponents approach the exact values for two-dimensional Ising model for lower learning rates.
 
\end{abstract}

\maketitle
\section{Introduction \label{Introduction}}

\begin{figure*}
    
    \includegraphics[width=0.9\linewidth]{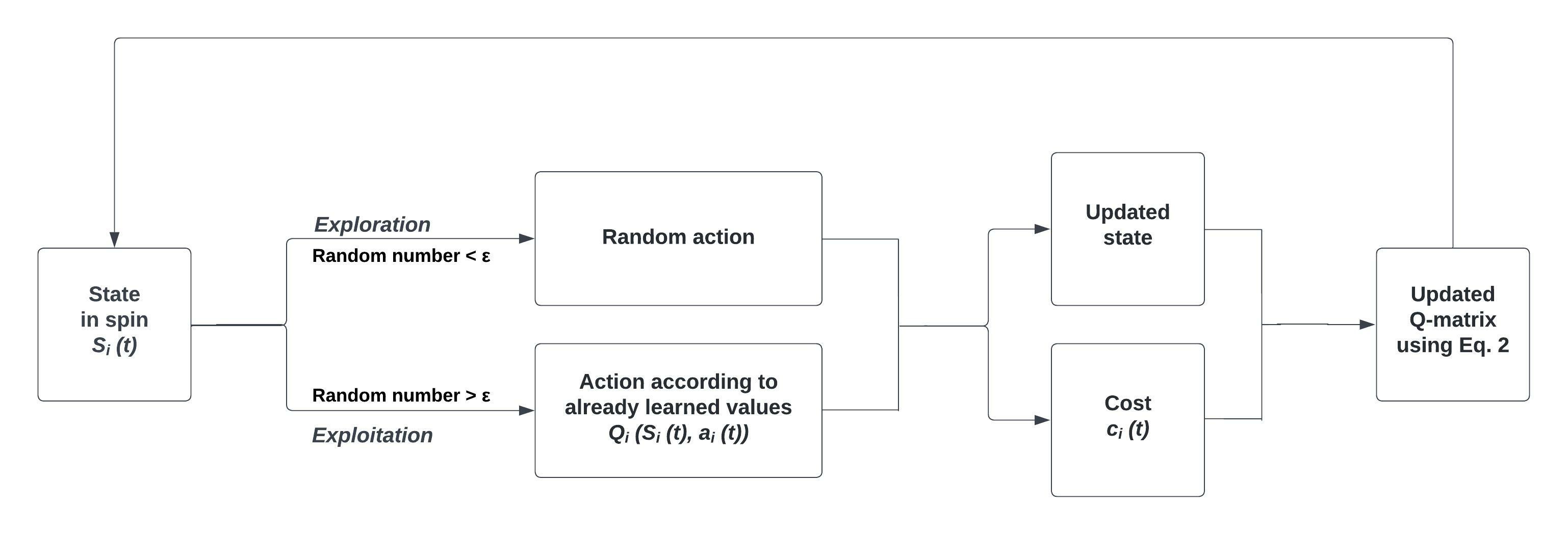} 
   
    \caption{Schematic block diagram of the model. The agent here is the spin with its own $\mathcal{Q}$-matrix. The state of the spin ($\mathcal{S}_i(t)$) is defined by whether it is in majority or minority. The spin can be in either of the states. The action is taken according to the $\epsilon$-greedy algorithm, that balances both exploration and exploitation. During exploitation, the spin's action is chosen as the action with the minimum $\mathcal{Q}$-value. Whereas in exploration, spin can either be flipped or left unchanged without any bias. Based on the action, updated state and cost are provided. $\mathcal{Q}$-matrix is updated according to Eq.\ref{qmatrixeq}. The spin uses the updated state and $\mathcal{Q}$-matrix and goes through the same process in a loop until the steady state is reached.}
 
    \label{fig:fig1}
\end{figure*}
The two-dimensional Ising  model \cite{Ising} is a prototype model which has been used to study various magnetic systems. The model has proven to be helpful in understanding the basic features of magnetic materials. Non-magnetic systems like the lattice gas model for liquid gas phase transition \cite{pathria, khuang} can also be  mapped to the Ising model. Several studies have investigated the Ising model theoretically and numerically \cite{onsager, binder1, binder2, binder3}. Prior numerical investigations of the model have predominantly been done using Maxwell-Boltzmann statistics and the Metropolis-Hastings algorithm. However, the system can be studied alternatively: where a spin learns from its experience and also from  its neighbours. Although in previous studies, the Ising model with memory is explored with some time-dependent coupling (Kernel) in the Hamiltonian \cite{memoryising1, memoryising2}. These models can be solved only for  specific kernels. This study aims to investigate a many-body spin system where the spin acts as an agent that learns from its experience. 
\par{}
In the current study, we develop an algorithm for the two-dimensional Ising spin system using a reinforcement learning (RL) approach \cite{RLbook, RL1, RL2}. RL is a branch of machine learning \cite{mlbook, mlpaper1, mlpaper2}. It is based on choosing suitable actions to maximize reward (appreciation) in a particular problem where the agent learns from its experience \cite{rlflocking, rlflocking1}. Previously, RL has been studied in many areas, such as game theory \cite{gametheory}, operations research \cite{opresearch}, information theory \cite{informationtheory},  statistics, etc. In recent years, an RL framework has been used to model interacting systems as well \cite{rlflocking, rlflocking1, rlflocking2, rlflocking3}. Our study uses RL to study the two-dimensional Ising spins. Here, the particles take action by considering the state of neighbours. To do so, we have used Q$-$Learning
\cite{mlpaper2}, which is a basic form of reinforcement learning where an agent uses Q$-$values (also called action values) to optimize its actions iteratively. The objective is to optimize a value function suited to a given environment. 
\par{}
We are introducing a way to study the many-particle interacting system, where we do not require a Hamiltonian, hence this can be applied to systems where the Hamiltonian is not known. The dynamics of the spins evolve through their states, actions (whether to change their orientation or not) and the reward associated with each action. The reward associated with the actions incentivizes the spin to align with the majority within its four nearest neighbours on the square lattice. At each step, spins learn from their previous actions and update the reward associated with their actions. This is done using the $\epsilon$-greedy algorithm \cite{greedy}, either to flip or retain its spin at each step. $\epsilon$ plays a role analogous to that of temperature in models which employ the Metropolis-Hastings algorithm to simulate the Ising model \cite{binder1}. The observables we are calculating are the measure of ordering and fluctuations present in the system, hence they can be compared with the physical observables like: magnetisation, susceptibility and the Binder cumulant. We find a transition from long-range ordered (finite magnetisation) to a disordered state (zero magnetisation) by tuning $\epsilon$ from small to large values. We studied systems of different sizes and learning rates, performed finite size scaling \cite{fss} and extracted the critical exponents $\beta$, $\gamma$ and $\nu$. 

The rest of the article is divided in the following manner. The next section \ref{model} gives the details of the  algorithm we introduced to study the system using reinforcement learning. In section \ref{results} we discuss our main results and calculation of various
observables. Section \ref{fss} discusses the finite size scaling, and finally, we conclude our study in section \ref{conclusion}.

\section{Learning model for two-dimensional Ising spin \label{model}}

We consider a two-dimensional system of Ising spins ($S=\pm 1$) on a square lattice of size $L \times L$ with periodic boundary condition in both directions. Previously, the Ising model 
is  exactly solved in one and two-dimensions or numerically studied using  the Metropolis-Hasting algorithm \cite{binder1}.  
In the Metropolis-Hastings algorithm, the temperature enters through the Boltzmann probability distribution $P \propto \exp(-\beta E)$, where $E$ is the local energy of the state and $\beta=\frac{1}{K_{B}T}$ is the inverse temperature \cite{khuang, pathria}. The ratio of interaction strength and temperature controls the degree of spin fluctuation. The system shows a transition from long-ranged ordered to disordered state in two dimensions on increasing temperature or decreasing interaction strength.

In this study, at each time step, we select a spin and define its state as $\mathcal{S}_i(t)=$ $+1$ if the spin is in the majority of its nearest neighbours (aligned with two or more nearest neighbour spins) and $\mathcal{S}_i(t)=$ $-1$ if it is in the minority (aligned with less than two nearest neigbour spins). Each spin has its own $\mathcal{Q}$-matrix corresponding to its states and actions ($\mathcal{Q}_i(\mathcal{S}_i(t), a_i(t))$). The  actions $a_i(t)$ of a spin are of two types: exploration (choosing a random action) and exploitation (choosing actions based on already learned $\mathcal{Q}$-values).

To prevent the action from always taking the same path and possibly over-fitting, we will introduce another parameter called $\epsilon$ to handle this during training. Instead of just selecting the optimal learned $\mathcal{Q}$-value action, the spin sometimes explores the action space further.  A higher $\epsilon$ value results in the spin taking actions with a greater cost on average. The action $a_i(t)$ is taken using the $\epsilon$-greedy algorithm \cite{greedy} for the parameter $\epsilon$ as given below:
\begin{equation}
    a_i(t)= 
\begin{cases}
	\it{argmin} \mathcal{Q}_i(\mathcal{S}_i(t),a_i(t)),& \text{probability } 1-\epsilon \\
   \text{random action},              & \text{probability } \epsilon 
\end{cases}
	\label{eqq}
\end{equation}
Thus, for large $\epsilon$ values, the
system performs more random exploration. We further define the cost function $c_i(t) =0 $ or $1$, if the chosen action leads the 
spin to its majority or minority respectively. The $\mathcal{Q}$-matrix is updated iteratively with the following equation as given 
in \cite{mlpaper2}
\begin{equation} \label{qmatrixeq}
\begin{aligned}
	\mathcal{Q}_i(\mathcal{S}_i(t), a_i(t)) \leftarrow \mathcal{Q}_i(\mathcal{S}_i(t),a_i(t))
     \\+ \alpha [ c_i({t}) - \mathcal{Q}_i(\mathcal{S}_i(t), a_i(t))]
\end{aligned}
\end{equation} 
where, $\alpha$ is the learning rate. We studied the system for different learning rates $\alpha=0.0001, 0.001, 0.05$, and $0.1$.  
The parameters $\epsilon$ and $\alpha$ control the exploration and exploitation respectively. A function of $(\epsilon, \alpha)$ will control the amount of fluctuations present in the spin degrees of freedom. For each realization, we  fixed $\alpha$  and varied $\epsilon$ as a temperature-like parameter in MCMC.   
The details of the RL scheme is explained in Fig. \ref{fig:fig1}. The scheme given in Fig. \ref{fig:fig1}  is repeated 
for all the spins one-by-one, and a time-step is defined as an update of all the spins at once. The total run time of the simulation is $t=1.2 \times 10^7$, and averaging is performed after simulation step of  $\tau=1.18 \times 10^7$. The system is studied for various sizes from $L=16$ to $96$. A total of $80$ different realisations are produced, and averaging is performed to improve the quality of data.
Each realization was initiated with a random distribution of up and down spins on the two-dimensional square lattice, and all the values of the $\mathcal{Q}$-values were set as $0$. The system is studied for various values of randomness parameter $\epsilon \in (0,1)$.

\section{Results\label{results}}

\begin{figure*}
\begin{subfigure}
  \centering
  \includegraphics[width=0.4\linewidth]{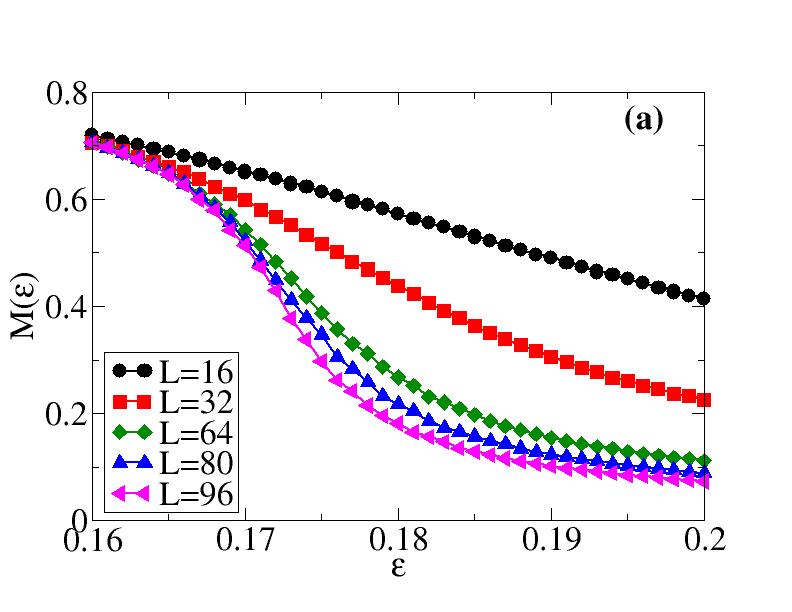}
  \label{fig:2a}
\end{subfigure}%
\begin{subfigure}
  \centering
  \includegraphics[width=0.4\linewidth]{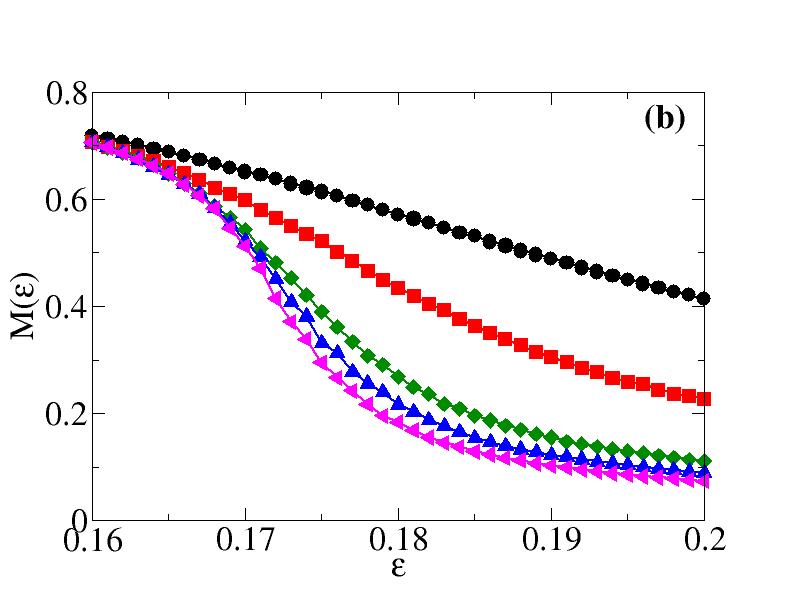}
  \label{fig:2b}
\end{subfigure}
\begin{subfigure}
  \centering
  \includegraphics[width=0.4\linewidth]{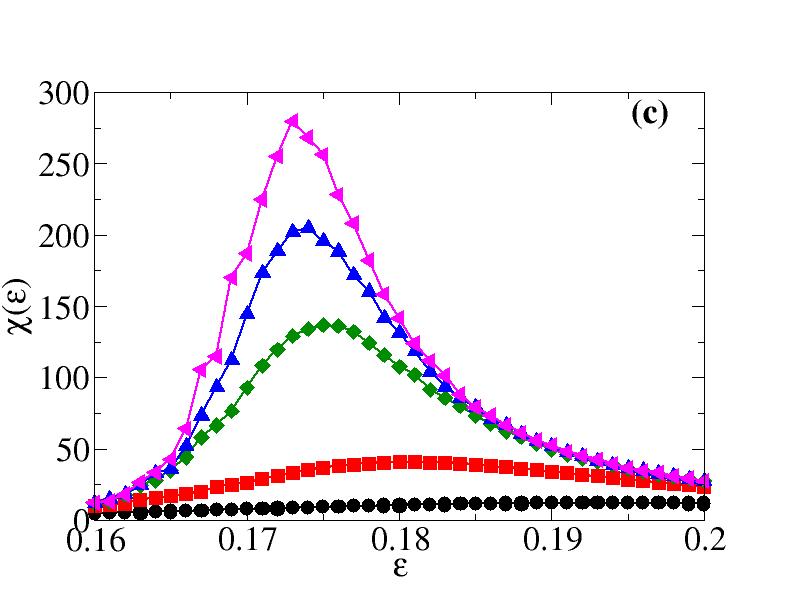}
  \label{fig:1c}
\end{subfigure}%
\begin{subfigure}
  \centering
  \includegraphics[width=0.4\linewidth]{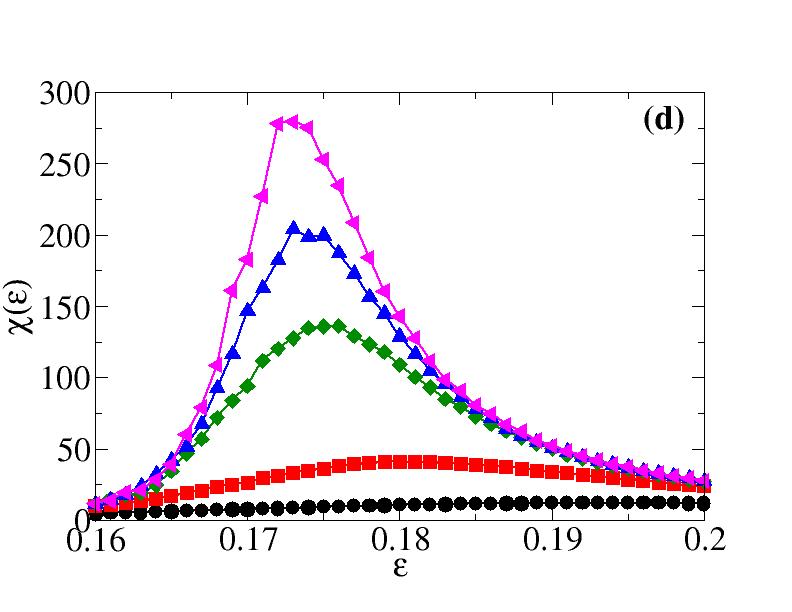}
  \label{fig:2d}
\end{subfigure}%
\begin{subfigure}
  \centering
  \includegraphics[width=0.4\linewidth]{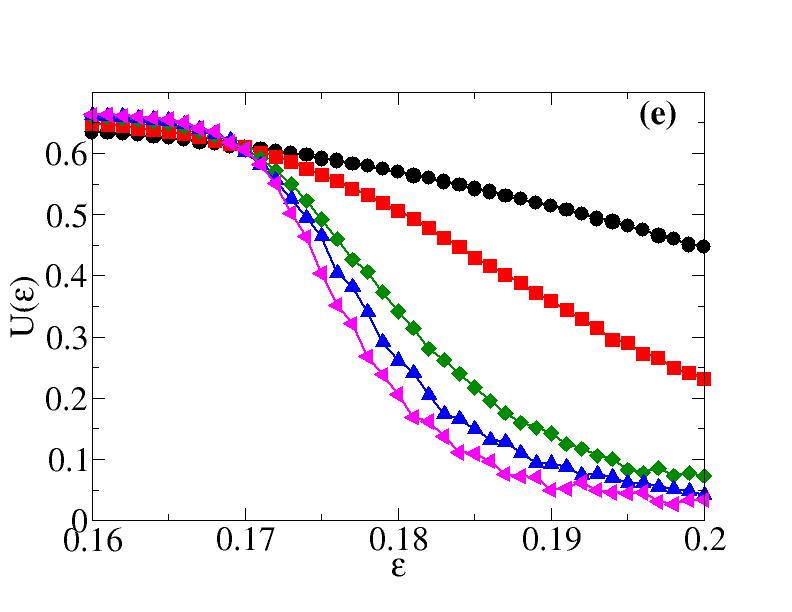}
  \label{fig:1e}
\end{subfigure}%
\begin{subfigure}
  \centering
  \includegraphics[width=0.4\linewidth]{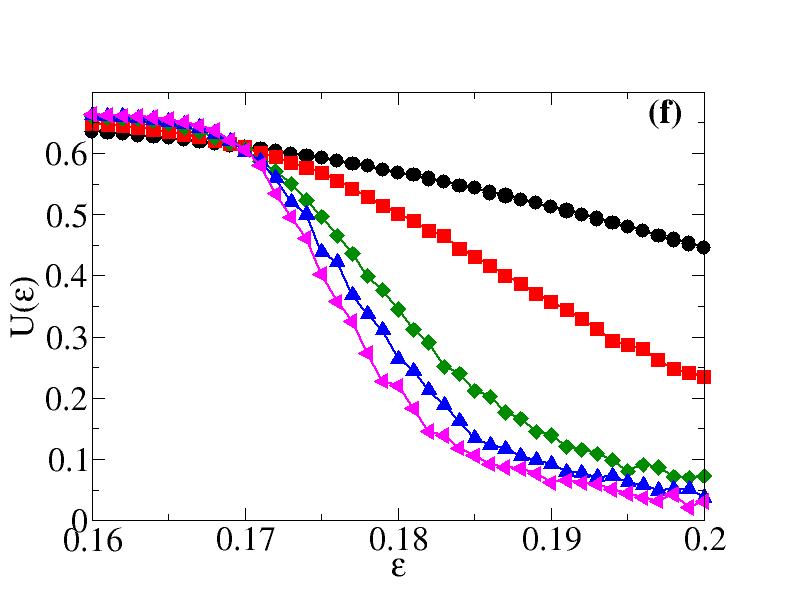}
  \label{fig:8f}
\end{subfigure}%
\caption{Variation of (a-b) $M$, (c-d) $\chi$ and (e-f) $U$ vs $\epsilon$ for $\alpha=0.0001$ and $\alpha=0.05$ respectively.}
\label{fig:fig2}
\end{figure*}
We define the magnetisation
order parameter $M(\epsilon)=<\frac{|\sum_{i=1}^{N}S_i(t)|}{N}>$, for different $\epsilon$, where $<..>$ implies the average value over time in the steady state and across multiple realizations. We show the plots for only two values of $\alpha=0.0001$ and $0.05$. Result (data) for other values of $\alpha$ is given in table \ref{table2}. 

In Fig. \ref{fig:fig2}(a-b), we plot the 
$M(\epsilon)$ vs. $\epsilon$ for different system size $L$ for $\alpha=0.0001$ and $0.05$ respectively. For all $L$, $M(\epsilon)$ 
remains close to $1$ (majority of spins ordered in the same direction) for small values of $\epsilon$ 
and approaches zero (random arrangement of spins on the lattice) as $\epsilon$ is increased. Thus, 
there is a phase transition from ordered $M \simeq 1$ to disordered state as a function of $\epsilon$. 
We further calculate the fluctuations in $M$, or the susceptibility, defined as 
$\chi(\epsilon) = <M(\epsilon)^2> - <M(\epsilon)>^2$, for different $L$, where $<..>$ have the same meaning 
as defined previously. The plot of $\chi(\epsilon)$ vs. $\epsilon$ is shown in Fig. 
\ref{fig:fig2}(c-d), for different system sizes $L$. We determine  $\epsilon_c(L)$ by calculating the location of maximum of susceptibility,  
$\chi_{max}(L)$. $\epsilon_c(L)$ decreases to lower $\epsilon$ for larger $L$.
We also calculated $\chi_{max}(L)$ for different system sizes and found that it increases with increasing $L$. To further characterize the phase transition, we calculate the fourth order moment of the mean magnetisation $M(\epsilon)$, the Binder Cumulant $(BC)$, defined as $U(\epsilon)=1-\frac{<M(\epsilon)^4>}{3<M(\epsilon)^2>^2}$. 
In Fig. \ref{fig:fig2}(e-f), we show the variation of $U(\epsilon)$ vs $\epsilon$ 
for different $L$. $U(\epsilon)$ remains close to $2/3$ in the ordered state for small $\epsilon$ and 
smoothly decays to zero in the disordered phase for large $\epsilon$. 
 We then performed finite size analysis of the data to characterize the critical behaviour in the thermodynamic limit, from the finite size $(L)$ data shown in Figs. \ref{fig:fig2}(a-b), \ref{fig:fig2}(c-d) and \ref{fig:fig2}(e-f).

\section{Finite size analysis \label{fss}}

\begin{figure*}
\begin{subfigure}
  \centering
  \includegraphics[width=0.4\linewidth]{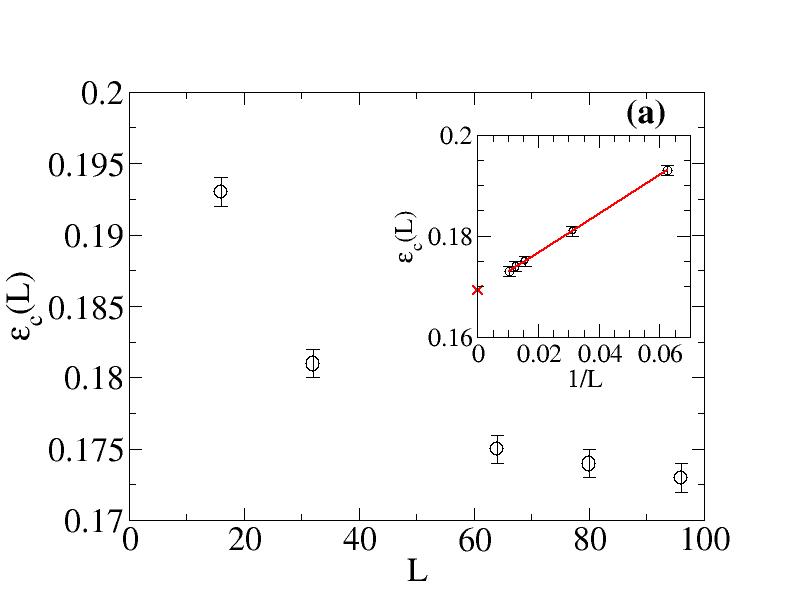}
  \label{fig:3a}
\end{subfigure}%
\begin{subfigure}
  \centering
  \includegraphics[width=0.4\linewidth]{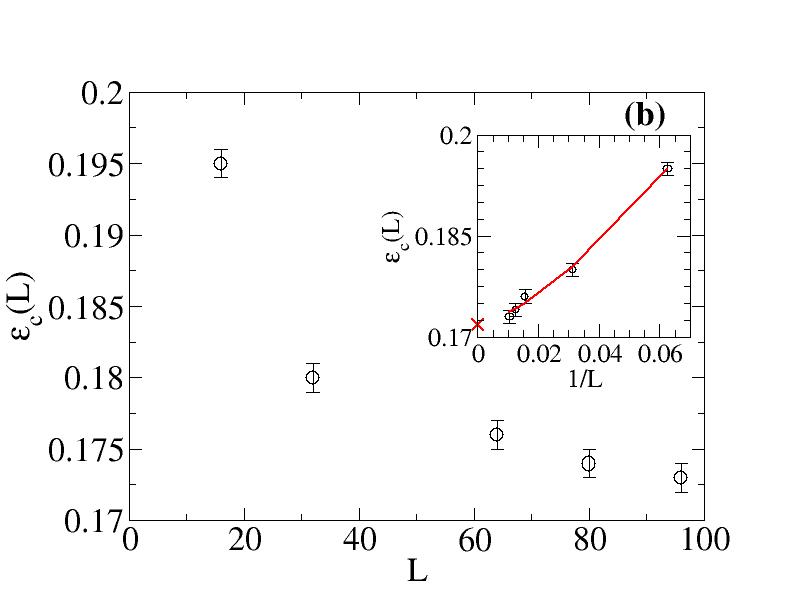}
  \label{fig:3b}
\end{subfigure}
\begin{subfigure}
  \centering
  \includegraphics[width=0.4\linewidth]{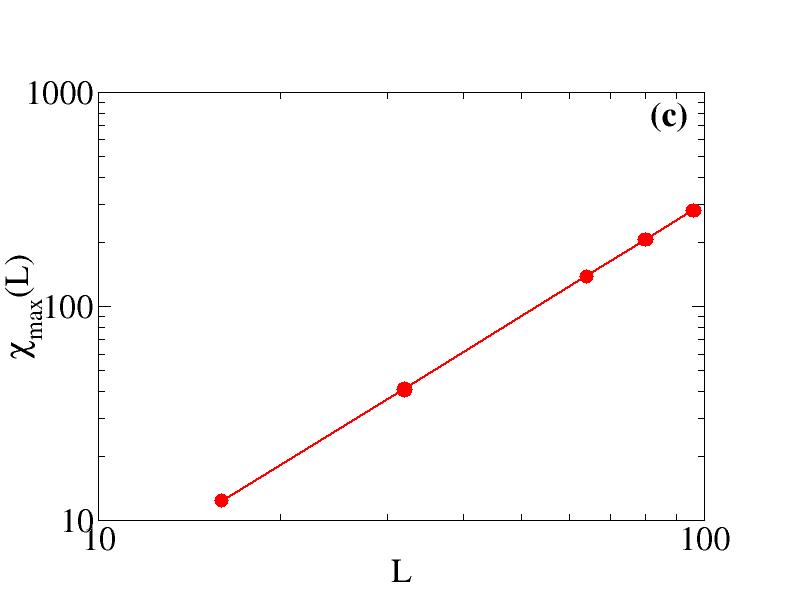}
  \label{fig:3c}
\end{subfigure}%
\begin{subfigure}
  \centering
  \includegraphics[width=0.4\linewidth]{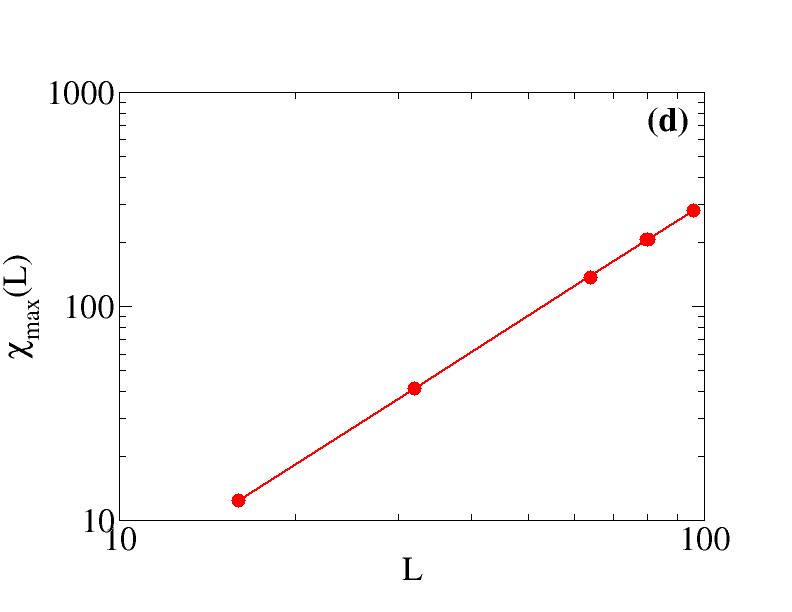}
  \label{fig:3d}
\end{subfigure}%
\caption{Variations of (a-b) $\epsilon_{c}(L)$ vs L, with insets showing the best fit to Eq. \ref{Eq4} and $\epsilon_c$ (marked as red cross on y-axis). [$\epsilon_c=0.1692\pm0.001$, $\nu=0.9894\pm0.00042$] and [$\epsilon_c=0.1718\pm0.001$, $\nu=0.6958\pm0.00153$] for $\alpha=0.0001$ and $\alpha=0.05$ respectively; and, (c-d) $\chi_{max}(L)$ vs L on a log-log scale. The lines in each are the best fits to the form $\sim L^{\gamma/\nu}$, where $\gamma=1.7253\pm0.01679$ and $1.2117\pm0.00058$ for $\alpha=0.0001$ and $0.05$ respectively.}
\label{fig:fig3}
\end{figure*}

\begin{table*}
    
\caption{}
\begin{tabular}{|c|c|c|c|c|c|c|}
\hline
$\alpha$  &$\epsilon_c$    & $\nu$ &   $\gamma$ & $\beta$ & $d\nu$ & $2\beta+\gamma$ \\ \hline
$0.0001$    &   $0.1692\pm0.001$    & $0.9894\pm0.00042$&  $1.7253\pm0.01679$   & $0.1127\pm0.00456$ & $1.9788\pm0.00084$ & $1.9507\pm0.02592$    \\ \hline
$0.001$  &     $0.1689\pm0.001$  &$0.9499\pm0.00167$ & $1.6562\pm0.02982$ &$0.1003\pm0.00456$& $1.8998\pm0.00343$ & $1.8568\pm0.06334$ \\ \hline
$0.05$ & $0.1718\pm0.001$ & $0.6958\pm0.00153$   &$1.2117\pm0.00058$ & $0.1167\pm0.01437$  &$1.3916\pm0.00306$ &$1.4451\pm0.0310$\\ \hline
$0.1$ &$0.17095\pm0.001$ & $0.7934\pm0.00315$ & $1.3739\pm0.00494$ & $0.1193\pm0.01227$&$1.5869\pm0.00632$ &$1.6125\pm0.02949$ \\ \hline
\end{tabular}
 
    \label{table2}
\end{table*}

\begin{table*}{hbt}
    
\caption{}
\begin{tabular}{|c|c|c|}
\hline
Critical Exponents    &  Two-dimensional Ising model theory \cite{3,13,11,16} \\ \hline
$\nu$    &   $1$     \\ \hline
$\beta$  &     $0.125$ \\ \hline
$\gamma$ &  $1.75$  \\ \hline
\end{tabular}
 
    \label{table}
\end{table*}

\begin{figure*}{hbt}
\begin{subfigure}
  \centering
  \includegraphics[width=0.4\linewidth]{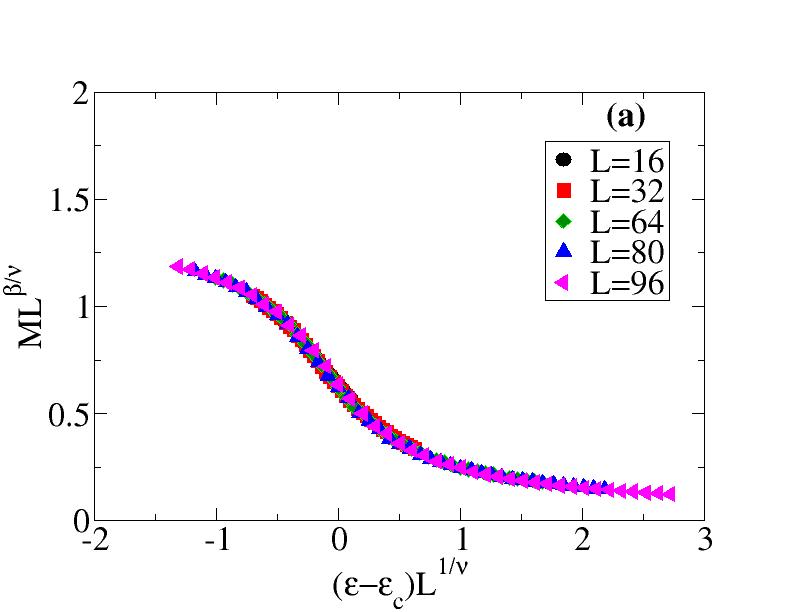}
  \label{fig:4a}
\end{subfigure}%
\begin{subfigure}
  \centering
  \includegraphics[width=0.4\linewidth]{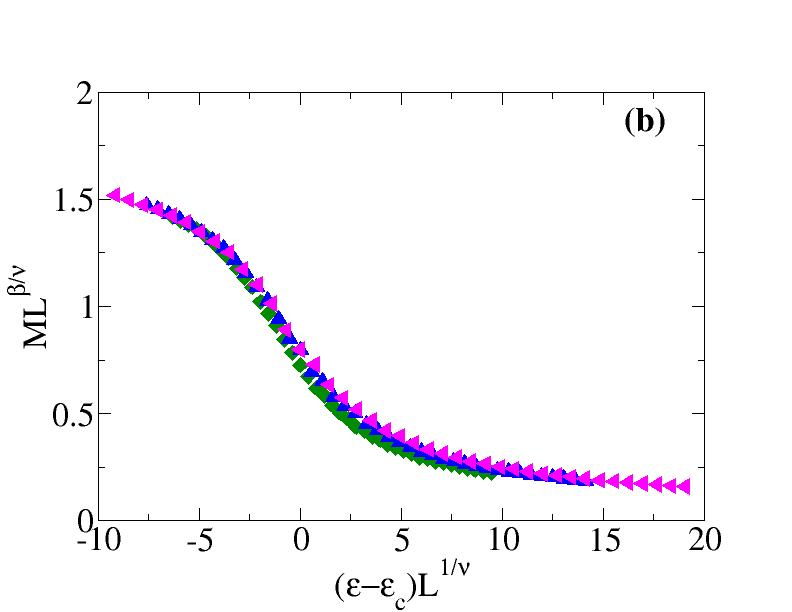}
  \label{fig:4b}
\end{subfigure}
\begin{subfigure}
  \centering
  \includegraphics[width=0.4\linewidth]{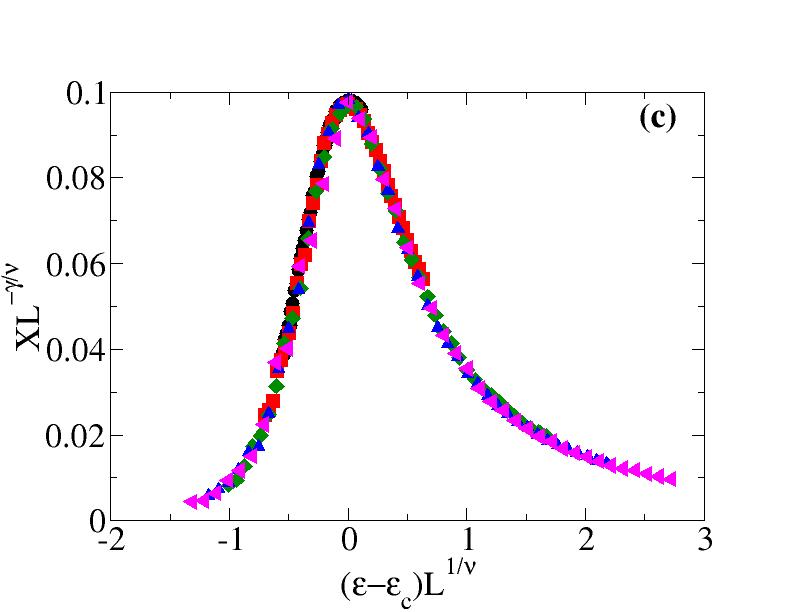}
  \label{fig:4c}
\end{subfigure}%
\begin{subfigure}
  \centering
  \includegraphics[width=0.4\linewidth]{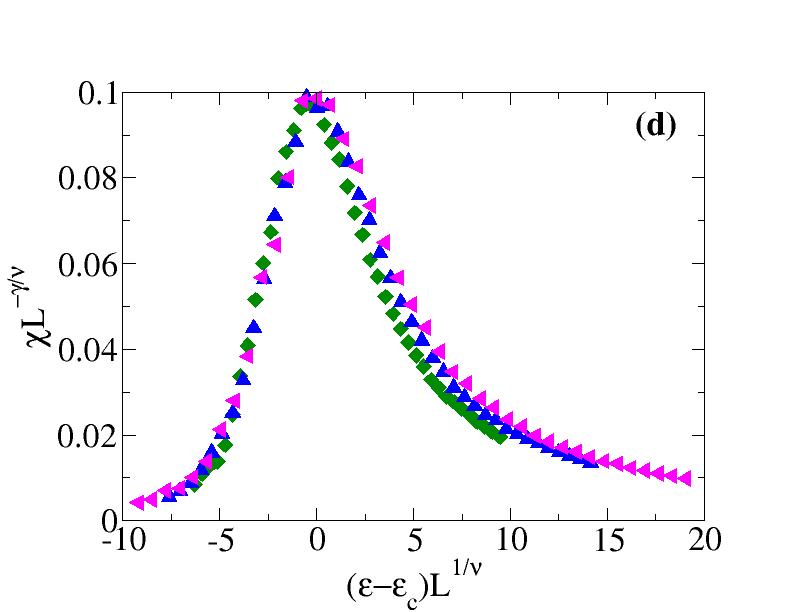}
  \label{fig:4d}
\end{subfigure}%
\begin{subfigure}
  \centering
  \includegraphics[width=0.4\linewidth]{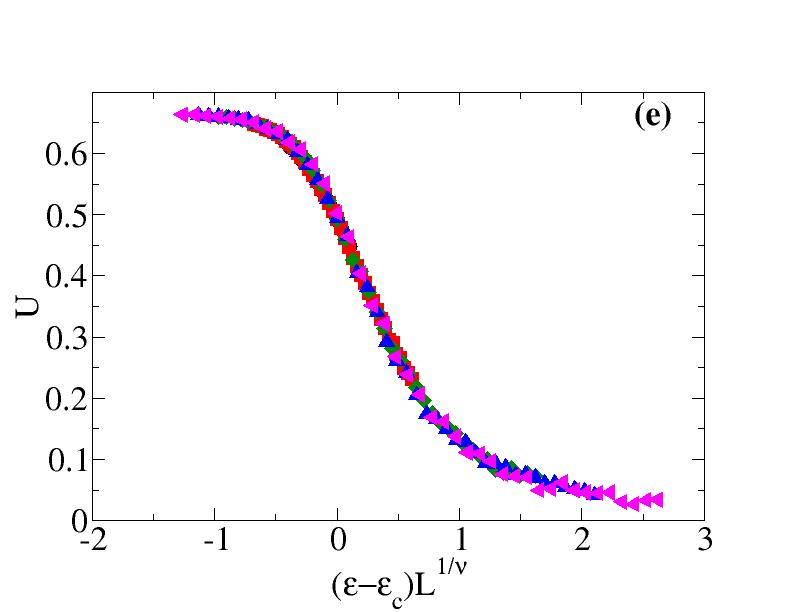}
  \label{fig:4e}
\end{subfigure}%
\begin{subfigure}
  \centering
  \includegraphics[width=0.4\linewidth]{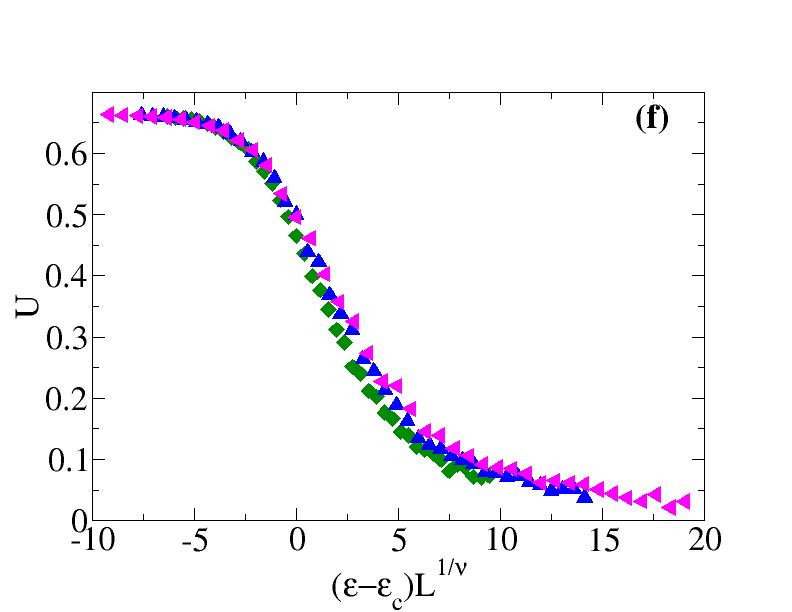}
  \label{fig:4f}
\end{subfigure}%
\caption{Scaling collapse of (a, b) $M(\epsilon)$, (c, d) $\chi(\epsilon)$ and (e, f) $U(\epsilon)$, by plotting according to the equations \ref{Eq3a}, \ref{Eq3b} and \ref{Eq3c} for $\alpha=0.0001$ and $\alpha=0.05$ respectively.}
\label{fig:fig4}
\end{figure*}

 We use finite size scaling (FSS) analysis to understand the dependence of the three quantities $M(\epsilon)$, 
 $\chi(\epsilon)$ and $U(\epsilon)$ on the system size $L \times L$. We assume that
the FSS forms for these quantities are the same as those for a
system of two-dimensional Ising model in previous studies \cite{binder1}.

\begin{subequations}
\label{eq3}
 \begin{align}
      M(\epsilon,L)=L^{-\beta/\nu}{\mathcal{M}}(\epsilon-\epsilon_c)L^{1/\nu}\label{Eq3a}\\
      \chi(\epsilon,L)=L^{\gamma/\nu}{\mathcal{\chi}}(\epsilon-\epsilon_c)L^{1/\nu}\label{Eq3b}\\
       U(\epsilon,L)={\mathcal{U}}(\epsilon-\epsilon_c)L^{1/\nu}\label{Eq3c}
 \end{align}
\end{subequations}

    where $\mathcal{M}$, $\mathcal{\chi}$  and $\mathcal{U}$ are the scaling functions for the 
    mean magnetisation,  susceptibility and the Binder cumulant  respectively. The exponents
    $\nu$, $\gamma$ and $\beta$ are the exponents for the correlation length, susceptibility and magnetisation 
    respectively for the standard Ising model \cite{fss}. 
    We are going to describe below the FSS analysis of our results within the numerical 
    accuracy of our data using RL model. We plot $\epsilon_c(L)$, the location of maximum of susceptibility $\chi_{max}(L)$ as a function of system size in 
    Fig. \ref{fig:fig3}(a-b) for $\alpha=0.0001$ and $0.05$ respectively. Further, we assume that the pseudocritical point for different system size $L$ is:
    \begin{equation}
       \epsilon_{c}(L)=\epsilon_c+c_1 L^{-1/\nu} \label{Eq4}
    \end{equation}
where, $\epsilon_c\equiv\epsilon_c(L \rightarrow \infty)$. In the insets of plots in Fig. \ref{fig:fig3}(a-b), the lines shows the fit of $\epsilon_{c}(L)$ with respect to $1/L$ and three parameters $\epsilon_c$, $c_1$ and $\nu$ are obtained from the fitting. The values of these three parameters we found are [$\epsilon_c=0.1692\pm0.001$, $c_1=0.3925$, $\nu=0.9894\pm0.00042$] and [$\epsilon_c=0.1718\pm0.001$, $c_1=1.2430$, $\nu=0.6958\pm0.00153$] for $\alpha=0.0001$ and $\alpha=0.05$ respectively. We further plot the $\chi_{max}(L)$ vs system size $L$ in the Fig. \ref{fig:fig3}(c-d) on $\log$-$\log$ scale. In a finite  system  of size $L$, $\chi_{max}(L) \sim L^{-z}$, where $z=\gamma/\nu$. This way, we extracted the exponent $\gamma$. The power law fit to the plot gives
$z=1.7438\pm0.0164$ and $z=1.7417\pm0.0047$ for $\alpha=0.0001$ and $\alpha=0.05$ respectively. 
Using the values of $\nu$ from Fig. \ref{fig:fig3}(a-b),  the exponent  $\gamma=1.7253\pm0.01679$ and $1.212\pm0.0006$ for $\alpha=0.0001$ and $0.05$ respectively. We further calculate the $\beta$ exponent for the magnetisation. At the critical $\epsilon_c$, we  calculate the root mean of magnetisation $M_{rms}= \sqrt{M{(\epsilon_c)}^2} \sim L^{-\beta/\nu}$ \cite{fss}.  
The power law fit provides the exponent $\beta/\nu$. We found that  $\beta=0.1127\pm0.0046$ and $0.1167\pm0.01437$ for the two values of learning rate $\alpha=0.0001$ and $0.05$ respectively.
Table \ref{table2}  shows the the values of the three exponents from our calculation for different $\alpha$ values  and the exact values of these exponents for the standard $2d$ Ising model are in given in Table \ref{table}. 
The exponents obtained for  $\alpha=0.05$ differ from the previous study of two-dimensional Ising model theory \cite{3,13,11,16}. But as we lower the learning rate $\alpha$, the exponents converge to the exact values for two-dimensional Ising model. 
Hence, slower learning rate converges the 
RL-model to the two-dimensional Ising model.
Altogether, our RL method is able to predict the  phase transition similar to as from the other methods and shows the similar phase transition characteristics. 

We further check the hyperscaling relation between the three exponents. The hyperscaling relation among the three exponents is $d \nu = 2 \beta + \gamma$, which is obtained from the scaling hypothesis of the underlying scaling functions, implies that all the observables can be calculated by the partition function of the system. $d=2$ is the dimensionality of space. We then substitute the values of the exponents we found in our RL study. The left hand side of the relation $d \nu$ and right hand side of the relation $2 \beta + \gamma$ is reported in sixth and seventh columns of Table \ref{table2}. 
We also observe that the accuracy of the hyperscaling relation improves on lowering learning rate $\alpha$. Hence again model converges to the two-dimensional Ising model for smaller $\alpha$.

We also show the plots of the scaling collapse of data for magnetisation $M$, susceptibility $\chi$ and Binder cumulant $U$ using the values of the three exponents found in our study for $\alpha=0.0001$ and $0.05$. In Fig. \ref{fig:fig4}(a-b), we plot the $ML^{\beta/ \nu}$ vs $(\epsilon-\epsilon_c)L^{1/ \nu}$  using Eq. \ref{Eq3a}, we find good collapse of data. We further plot the scaled susceptibility $\chi L^{-\gamma/ \nu}$ vs $(\epsilon -\epsilon_c(L))L^{1/ \nu}$ in Fig. \ref{fig:fig4}(c-d) using Eq. \ref{Eq3b}  and find good collapse of data for different system sizes. Similarly, we plot the $U$ vs. scaled $\epsilon$, $(\epsilon-\epsilon_c)L^{1/ \nu}$ in Fig. \ref{fig:fig4}(e-f) using Eq. \ref{Eq3c} and again find good scaling collapse of data.  Scaling collapse of data also improves by lowering the learning rate. We also find the scaling collapse for other $\alpha$ values (data not shown).    
\section{Discussion\label{conclusion}}
In this work, we used an RL framework to investigate Ising spins in two dimensions with periodic boundary condition in the both directions. 
Each spin can have two characteristic states and it can switch between these states using an action selected using the $\epsilon$-greedy algorithm. We find that if the spins get rewarded for staying in the majority then the system shows a phase transition from disordered to ordered state on decreasing $\epsilon$. Hence, $\epsilon$ plays a role similar to the temperature in the Metropolis Monte-Carlo for two-dimensional Ising model.
    
We further characterise the phase transition by calculating the critical point and different critical exponents for magnetisation, susceptibility and correlation length $\beta$, $\gamma$ and $\nu$ respectively.
Data shows scaling collapse for scaled magnetisation, susceptibility and Binder cumulant. 
We observe the exponents match with the exact exponents for two-dimensional Ising model for the lower learning rate $\alpha$. Hence our RL model converges to the two-dimensional Ising model for lower learning rates. 

Our current study provides a  reinforcement learning approach to study the spin system. It can  be further used for other many particle interacting systems, where the form of Hamiltonian is not known. The rudimentary nature of this approach makes it adaptable to study many particle complex systems \cite{rlflocking,rlflocking2,vicsek,pottsmodel}.


\section{Acknowledgement \label{Acknowledgement}}
SM thanks DST-SERB
India, ECR/2017/000659 for the partial financial support.
The
support and the resources provided by PARAM Shivay
Facility under the National Supercomputing Mission,
Government of India at the Indian Institute of Technology, Varanasi are gratefully acknowledged. Computing facility at Indian Institute of Technology(BHU),
Varanasi is gratefully acknowledged.
\bibliographystyle{apsrev4-1}
\bibliography{references}


\end{document}